\documentclass{mem}
\usepackage{natbib}
\usepackage{txfonts}
\usepackage{balance}
\usepackage{graphicx}
\usepackage{flushend}
\usepackage[breaklinks,pdftex]{hyperref}
\idline{75}{282}
\begin{document}
\def\teff{$T\rm_{eff }$}
\def\kms{$\mathrm {km s}^{-1}$}

\newcommand{\gp}[1]{\textcolor{teal}{\textbf{GP:} #1}}

\title{
Planet Formation and Disk Chemistry
}

   \subtitle{Dust and gas evolution during planet formation}
   
\author{
G. Perotti\inst{1} 
\and L. Cacciapuoti\inst{2} \and N.-D. Tung\inst{3} \and T. Grassi\inst{4,5} \and E. Schisano\inst{6} \and L. Testi\inst{7,8}
}

\institute{
Max-Planck-Institut f\"{u}r Astronomie, K\"{o}nigstuhl 17, 69117 Heidelberg, Germany
\email{perotti@mpia.de}
\and
European Southern Observatory, Karl Schwarzschild-stra\ss e 2, D, 85748 Garching bei M\"{u}nchen, Germany
 \and
Université Paris-Saclay, Université Paris Cité, CEA, CNRS, AIM, 91191, Gif-sur-Yvette, France
 \and
 Max-Planck-Institut f\"ur Extraterrestrische Physik, Giessenbachstra\ss e 1, 85748 Garching bei M\"{u}nchen, Germany 
 \and
 Exzellenzcluster ``Origins'', Boltzmannstr. 2, D-85748 Garching, Germany
\\
The remaining affiliations can be found at the end of the paper.
}

\authorrunning{Perotti et al.}

\titlerunning{Planet Formation and Disk Chemistry}

\date{Received: XX-XX-XXXX; Accepted: XX-XX-XXXX}

\abstract{Over the past decade, progress in observational capabilities, combined with theoretical advancements, have transformed our comprehension of the physics and chemistry during planet formation. Despite these important steps forward, open questions persist on the chemical and physical evolution of solids in their journey from the collapsing molecular cores to disks and planetary bodies. This chapter is a repository of such burning questions. It has the ambition to identify the most promising avenues for future research based on current observational and modeling opportunities. 

}
\maketitle{}

\section{State of the art} 
\begin{figure*}
    \centering
    \includegraphics[width=\hsize]{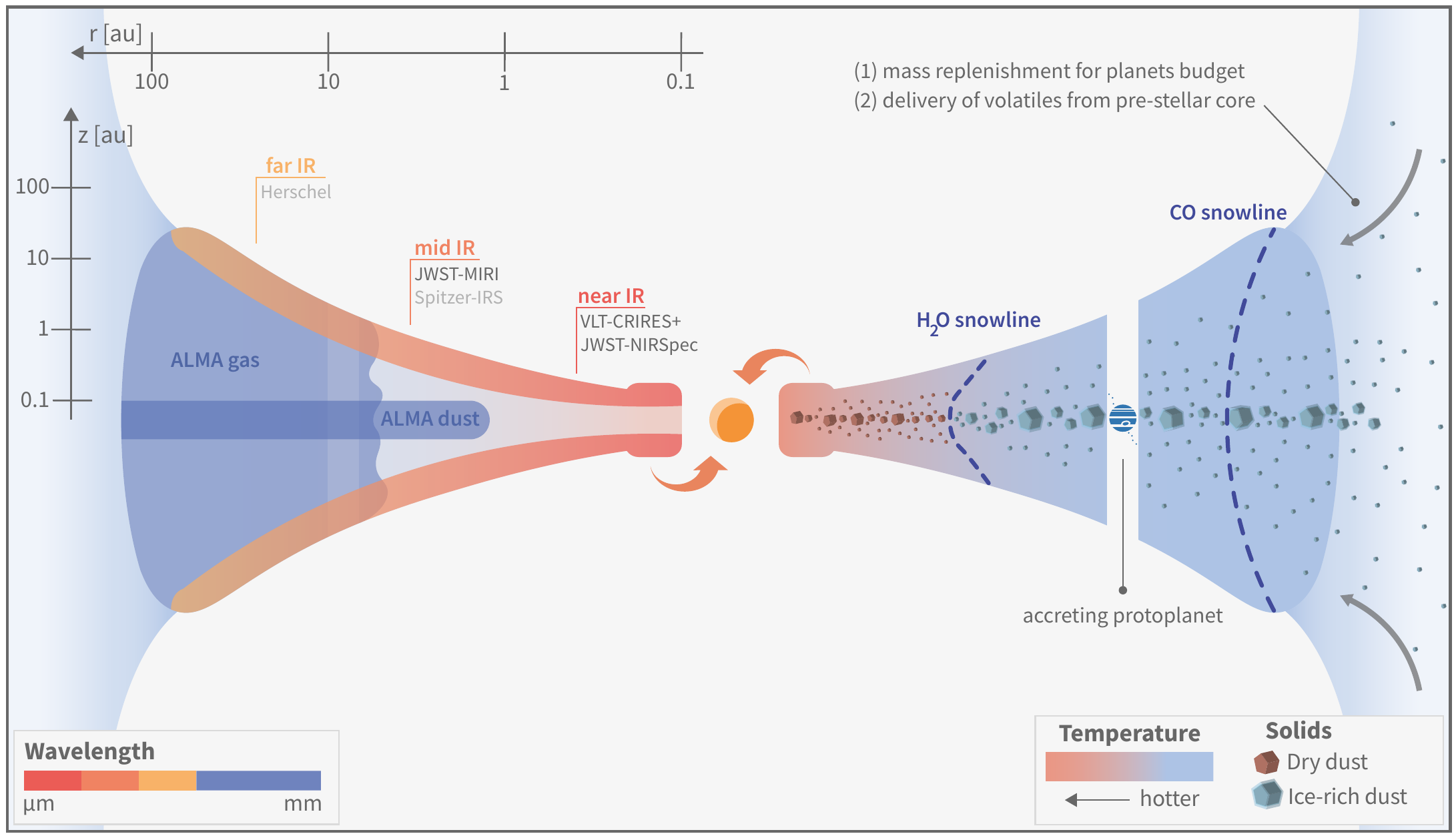}
     \caption{Schematics of a planet-forming disk. The illustration highlights facilities used to probe different dust and gas disk reservoirs. Figure by A. Houge.}
    \label{fig:disk_sketch}
\end{figure*}

Protoplanetary disks (PPDs; Figure~\ref{fig:disk_sketch}) were first imaged at high spatial resolution about a decade ago \citep[e.g.,][]{2015ApJ...808L...3A}. Since then, two major findings have been impacting this growing field of research. The first one is that dust substructures are ubiquitous in disks and they can be the signpost of accreating planets \citep[e.g.,][]{Andrews2020}. The second one is that planet formation starts early, already during the embedded phase of star formation, a few 10$^5$ yrs after core collapse, as deduced from the evolution of disk masses in comparison to exoplanet populations \citep[e.g.,][]{2018A&A...618L...3M,2019ApJ...875L...9W,Testi2022}, implying that star and planet formation are not two distinct processes on a temporal sequence, but one. This paradigm shift evidences the importance of accounting for the large-scale environment on disk evolution and in shaping the properties of planet-building material \citep[e.g.,][]{Pineda2023}. 

For editorial reasons, in this chapter, we will uniquely explore the following topics. Section~\ref{dust} addresses how dust properties evolve during the journey from protostellar envelopes to disks to planets. Section~\ref{volatiles} discusses the current challenges to link the chemical properties of the gas in disks with the composition of exoplanetary atmospheres. Section~\ref{modelling} explores the need for more detailed numerical models to accurately interpret high-resolution observational data of current facilities. 

Given the limited space available, references are primarily to reviews; we encourage the reader to cite the original articles. In particular, we refer the reader to recent overviews on dust evolution \citep{Birnstiel2023} and disk chemistry \citep{Oeberg2023}.  

\section{Evolution of dust properties: from envelope to disk}
\label{dust}
\begin{figure}
    \centering
    \includegraphics[width=0.75\linewidth]{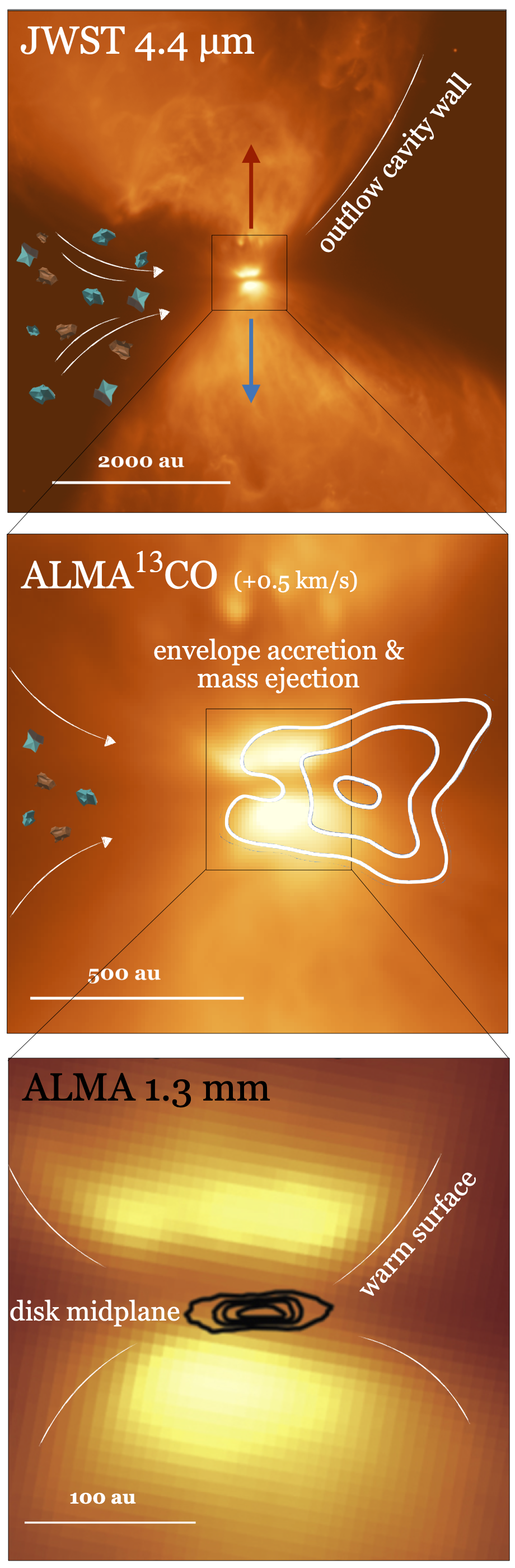}
     \caption{Complementary JWST and ALMA observations the L1527 IRS protostar. \textit{Top:} JWST/NIRCam image highlighting the wide molecular outflow (PID: 2739). \textit{Middle:} ALMA $^{13}$CO contours showing a snapshot of gas moving at +0.5 km/s with respect to the central source (PID: 2019.1.00261.L). \textit{Bottom:} ALMA 1.3~mm dust emission contours probing the disk midplane (adapted from \citealt{2023ApJ...951...10V}).}
    \label{fig:L1527}
\end{figure}

Protostellar envelopes and ``streamers’’, whose molecular gas properties and kinematics were covered in the previous chapter, are the ducts through which protoplanetary disks are replenished with material from the interstellar medium (Figure~\ref{fig:L1527}). The dust transported within these infalling structures also plays critical roles in the evolution of the inner systems.

These infalling reservoirs of material can significantly increase the mass budget for planet formation (e.g., \citealt{Pineda2023}, \citealt{2024A&A...682A..61C}), they deliver volatiles onto the disk via dust grains, whose surfaces are favorable sites for the formation of complex molecules in the ISM \citep{Tielens1982}. Moreover, envelope grains couple with the protostar’s magnetic field and contribute to the transport of angular momentum via magnetic breaking, thus contributing to regulate the sizes of rotationally supported disks (e.g., \citealt{Zhao2016}). Finally, infalling material can induce shocks (e.g., \citealt{Garufi2022}) and instabilities in the disks, that can even lead to the formation of the first generation of planetary cores (e.g., \citealt{Cridland2022}). It is thus important to characterize dust properties while bearing in mind the complex and intertwined picture of star and planet formation, a process happening on different scales, densities, temperatures and timescales (Figure~\ref{fig:disk_sketch}).

\begin{figure}
    \centering
    \includegraphics[width=\linewidth]{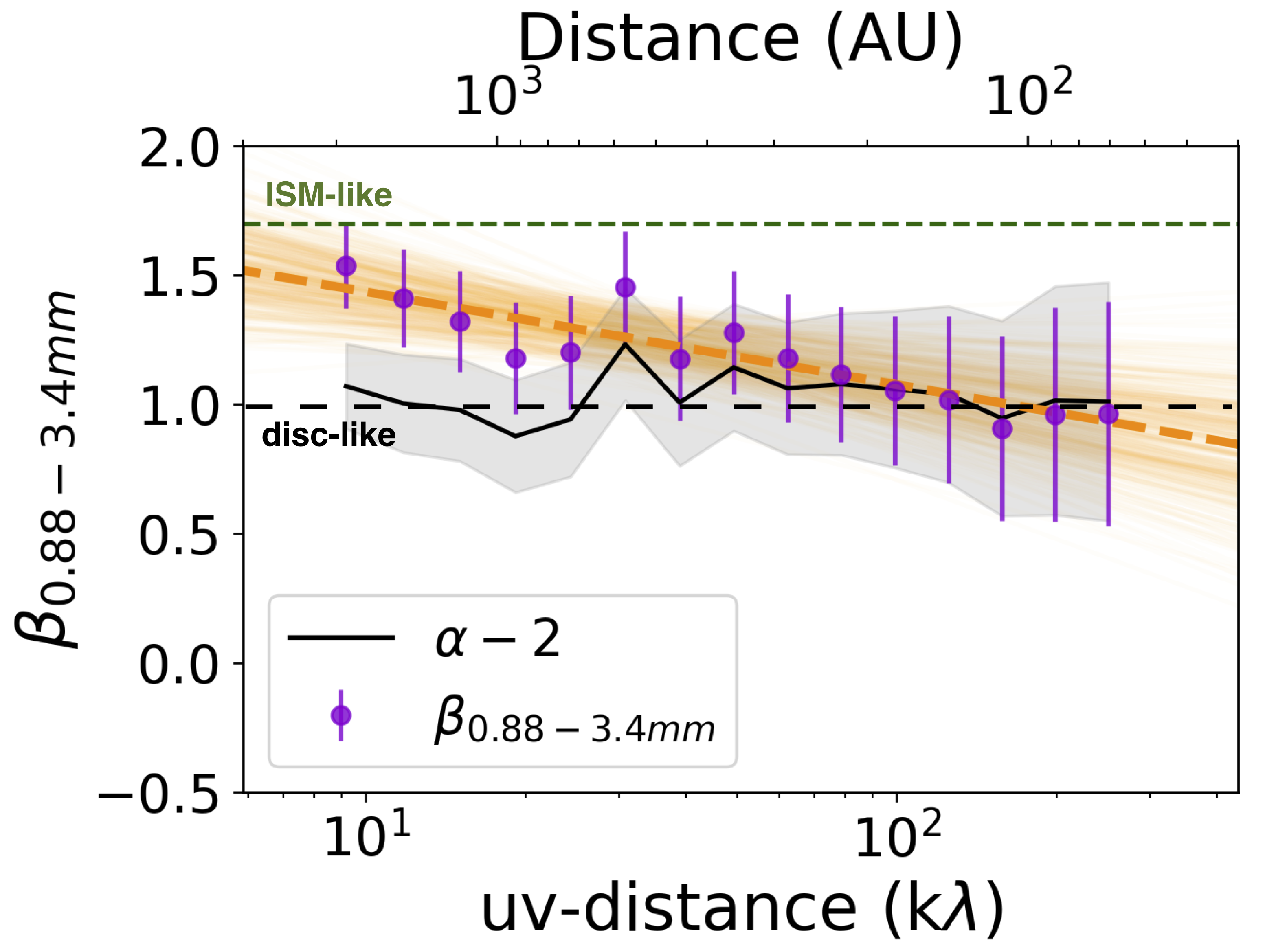}
    \caption{Multi-wavelength dust opacity spectral index ($\beta_{0.88-3.4\,mm}$) across the envelope of L1527, as a function of ALMA antennas baseline length or, equivalently, recovered physical scale (top x-axis). The probed scales go from 2000~au to 100~au. A gradient in dust properties across scales could be explained by dust growth to sub-millimeter sizes, variations in the chemical compositions of grains, or outwards transport of disk grains along outflows. Adapted from \citep{Cacciapuoti2023}.}
    \label{fig:LC23}
\end{figure}
Recent works have shown that dust properties gradually change from outer (10$^{3-4}$ au) to inner (10$^{2-3}$ au) envelope (\citealt{Galametz2019}, \citealt{Cacciapuoti2023}; Figure~\ref{fig:LC23}). Dust polarized thermal emission supports these findings. When aligned to protostellar magnetic fields, dust grains emit thermally with a preferential polarization, and the fraction of polarized light depends on the maximum grain sizes of the dust distribution \citep{Valdivia2019}. The polarization fractions so far measured in protostellar envelopes (e.g., \citealt{Galametz2018}) are consistent with grains of up to a few tens of microns, much larger than the sub-micron grains of the diffuse ISM. However, analytical and numerical simulations of collapsing cores seem to show limited grain growth of only up to a few microns in envelopes (e.g., \citealt{Testi2014}, \citealt{Silsbee2022}, \citealt{Lebreuilly2023}). To solve this tension, we will need more robust interpretations of dust observations by refining dust models (e.g., \citealt{Ysard2019}), and to explore alternative explanations to the observations (as dust lifting from disks to envelopes, \citealt{Tsukamoto2021}, \citealt{Cacciapuoti2024}).  Simulations should also account for often disregarded effects like the back reaction of dust on gas and the charge of dust grains (e.g., \citealt{Hennebelle2023}).

When dust reaches the disk, further growth into cores happens. This might occur earlier than expected in disks. The evolution of the dust content in disks is far from being clear as solids are  converted into unobservable large bodies and uncertainties on disk opacity at young ages imply large uncertainties in the dust mass measurements at ages $\le$0.5~Myr (e.g. \citealt{2019ApJ...875L...9W}, \citealt{Testi2022}, \citealt{2024A&A...684A..36T}).
Early planetesimals formation might also justify disk substructures that have been observed already in young disks (e.g., \citealt{Seguracox2020}) and the insurgence of a secondary dust distribution generated by giant impacts \citep{2012ApJ...750....8T,Testi2022,Bernabo2022}.

Observational studies are now aiming to further constrain dust properties in star and planet-forming environments:
\begin{itemize}
    \item Are dust grains evolving beyond expectations at envelope scales?
    \item How much mass do they deliver to the inner system?
    \item Which volatile species do they transport in the planet-forming disk?
\end{itemize}
Current and future studies will have to take into consideration the key inherent degeneracy of grain sizes and geometry, composition, and porosity, when trying to characterize the dust content of star and planet-forming environments. Different observatories will have to target these regions at several wavelengths. In this direction, JWST programs are targeting young sources in the near- and mid-infrared regime to tightly constrain the bulk and ice compositions of dust grains across scales \citep{mcclure2023,Sturm2023} and eventually into exoplanets atmospheres. Finally, sensitive radio interferometers are starting to exploit the power of dust thermal polarized emission observations in disks, a means to learn about dust sizes and porosity \citep{Zhang2023,Lin2024} in an independent fashion. Deep observations will be needed to draw similar, more accurate identikits of dust in fainter, collapsing envelopes (e.g., \citealt{Maury2022}).

\subsection{Dust mass budget for planet formation}
As discussed in Section~\ref{dust}, there is general consensus that the assembly of planetary embryos and cores has to at least start at the Class~0/I protostellar stages. 
Even if disk mass measurements in the Class 0/I phase are very uncertain, very controversial evidence for a gradual decrease of disk mass from Class~0, I, to II YSOs in Perseus has been presented \citep{Tychoniec2020}. 
Similar evidence had been previously claimed for Taurus (\citealt{2014MNRAS.445.3315N}), Orion (\citealt{2020ApJ...890..130T}) and Ophiuchus (\citealt{2019ApJ...875L...9W}), although the estimates in the different regions vary by orders of magnitude, reflecting the inherent difficulties in providing robust estimates \citep[see also][for a discussion on the uncertainties in measuring protostellar disk masses]{2024A&A...684A..36T}.

Even though direct measurements of the disk mass in Class~II exist (in terms of gas mass and gas-to-dust ratio) from the gas component such as using hydrogen deuteride \citep{2013Natur.493..644B} or the dynamical non-Keplerianity induced by self-gravity (e.g., \citealt{2021ApJ...914L..27V,2024arXiv240212236M}), dust grains are still the most accessible and abundant probes of the disk material. Thus, before continuing the discussion on the disk's mass budget, let us take one step back and review the measurements of the disk mass using dust as the tracer. 

\begin{figure}
    \centering
    \includegraphics[width=0.5\textwidth]{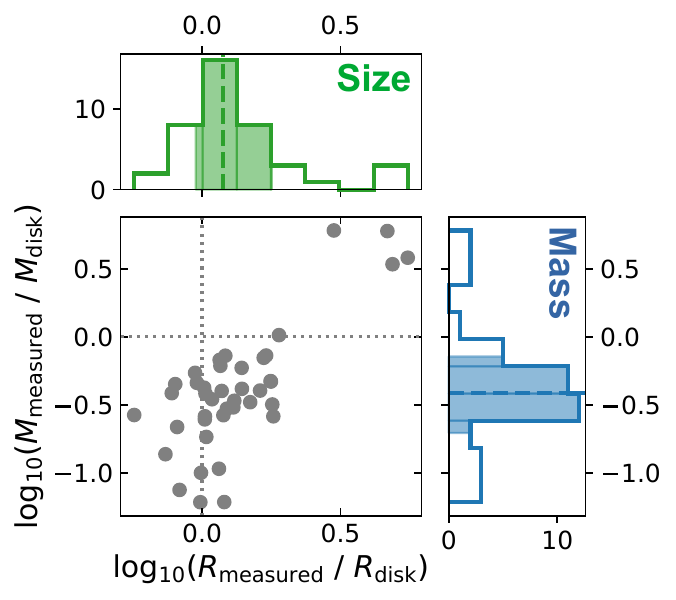}
    \caption{Radii and masses measured from the synthetic millimetric fluxes at $0.89~{\rm mm}$ for a population of disks formed self-consistently in one MHD simulation by \citet{2024A&A...682A..30L}. The disk sizes can be retrieved with good accuracy within an uncertainty of factor $\sim 2$, whereas the masses are erroneously estimated due to high optical depth. Adapted from \cite{2024A&A...684A..36T}.}
    \label{fig:size_mass}
\end{figure}

A fair amount of studies of Class 0/I disks measure disk masses from the modeling of the dust emission - either on the image plane or in the $uv$ Fourier space - and still rely on simple analytical prescriptions, such as the empirical Gaussian (e.g., \citealt{2020A&A...633A.114S}) or the Nuker profile (e.g., \citealt{2017ApJ...845...44T}), or for better separation of the disk from the envelope's emission, an additional component for the extended structure (\citealt{2019A&A...621A..76M}) needs to be included. Especially in young disks, studies that do not include a subtraction of the large scale emission may result in incorrect estimates not only for the disk mass, but also the disk radius, hence the results cannot be trusted \citep{2024A&A...684A..36T}. The mass is then determined by the optically thin conversion from the total fluxes assuming certain values for the (often average) temperature and the dust opacity. However, these have been proved to wrongly estimate the mass at mm/sub-mm wavelengths up to at least $\sim1~{\rm mm}$ where the disk emission is known to be optically thick. Figure~\ref{fig:size_mass} shows a compilation of the measured disk radii and masses from the synthetic observations at $0.89~{\rm mm}$ performed on one of the simulations by \cite{2024A&A...682A..30L}, in which the disk sizes are accurately estimated with acceptable uncertainty but the masses cannot be measured under the optically thin assumption. ALMA observations at longer wavelength such as one achievable in Band 1 ($6-9~{\rm mm}$) are expected to help reducing the optical depth effect hindering us from probing the disk material. More robust modeling techniques, less prone to the optical thickness of the medium, such as the radiative-transfer-powered models used by \cite{2022ApJ...929...76S} or modern statistical-based tools such as machine learning, are other promising alternatives, although they often also have their own set of caveats and can sometimes be detached from the real physics. In addition, and especially in the case of young disks (Class 0 and I), the assumption of average temperatures similar to those of class~II disks is also an unrealistic assumption leading to large overestimates of the disk dust mass \citep{2024A&A...682A..30L}.

On the theory and numerical simulation side, we are currently in possession of state-of-the-art multi-scale MHD simulations (\citealt{Kuffmeier2017,2024A&A...682A..30L}) that enable us to follow the formation and evolution of both the gas and the dust disks (including, but not limited to, their masses), starting from the collapse of the molecular clouds, based on various unconstrained initial conditions. These simulations are designed to help us in the search for the answers to many long-standing puzzles regarding disk evolution and the formation of planets, such as:
\begin{itemize}
    \item What do we know of disk evolution? Can we constrain the role of viscous evolution versus MHD wind transport?
    \item What is the connection between substructures, planet formation, and the assembly of their atmospheres?
\end{itemize}
By means of realistic radiative transfer post-processing and synthetic observations (e.g., \citealt{2020ApJ...905..174A, 2024A&A...684A..36T}), we would be able to predict the multi-wavelength emission of the disks in these simulations and compare the inferred properties with observations using the same modeling techniques. Such comparisons would be the key to understanding the theories of the dust disk evolution and planet formation, especially the dust mass budget in different evolutionary stages.

\vspace{-0.3cm}
\section{Evolution of the gas-phase chemistry in disks}
\label{volatiles}

PPDs share the same bulk elemental composition of their central stars, but their chemical composition strongly varies across the radial and vertical structure due to the different temperature, density, and UV irradiation conditions present across the disk. 

Volatiles are simple molecules composed of oxygen (O), carbon (C), nitrogen (N), sulfur (S), and phosphorus (P), that have low sublimation temperatures, so they remain gaseous for extended regions in PPDs. ALMA and JWST are detecting a growing number of volatiles in disks \citep{Kamp2023,Oeberg2023}, with the most abundant being CO, H$_{2}$O, CO$_{2}$. Since each volatile freezes at a different temperature, the disk is composed of different regions, delimited by icelines, where specific molecules are depleted from the gas phase. This leads to a stratification in the chemical structure due to the efficiency with which different elements are locked up in ices. The position of the icelines, then, separate the disk into regions characterized by different values for the main elemental ratios, C/O, C/N, N/O.  

Most of the disk volatiles, CO, H$_{2}$O, CO$_{2}$, basic organics and few N-bearing species, like N$_{2}$ and NH$_{3}$, are formed in molecular clouds, where they are typically locked in the icy mantles on dust grains. It is rather unclear if the material undergoes a massive chemical reprocessing, altering its natal composition, during the formation of PPDs. This depends on the conditions to which the material is exposed while being transferred to the disk: if it crosses cold regions, shielded from direct UV radiation, such as the one delivered to the outer disk, it does not alter significantly its chemical composition \citealt{Aikawa2022,Bergin2023}. On the contrary, a substantial change in the chemical composition happens for material exposed to high temperature and strong UV fields, that dissociate molecules and shorten the chemical reaction timescale. This reset in composition is certainly active in the inner and high disk surface layers. Evidences of this process are found in the inner region of our Solar System \citep{Pontoppidan2014}. 

PPDs are highly dynamical, with their thermo-physical structure varying with time. The disk material is transferred both vertically and radially due to viscous accretion, turbulence, diffusion, radial drifting, settling and dust growth \citep{Birnstiel2023}. The local chemical composition is continuously shuffled by these processes, for example, the drift of icy grains leads to a replenishment of water vapor in the inner disk \citep{Banzatti2023,Perotti2023}. Moreover, young PPDs are exposed to numerous luminosity outbursts, due to sudden increases in stellar mass accretion. These events heat up the disk material and irradiate it, resulting in the sublimation of ices and activating chemical reactions, affecting the chemical composition of the disk material \citep{Houge2023}. 

\begin{figure}
    \centering
    \includegraphics[width=0.45\textwidth]{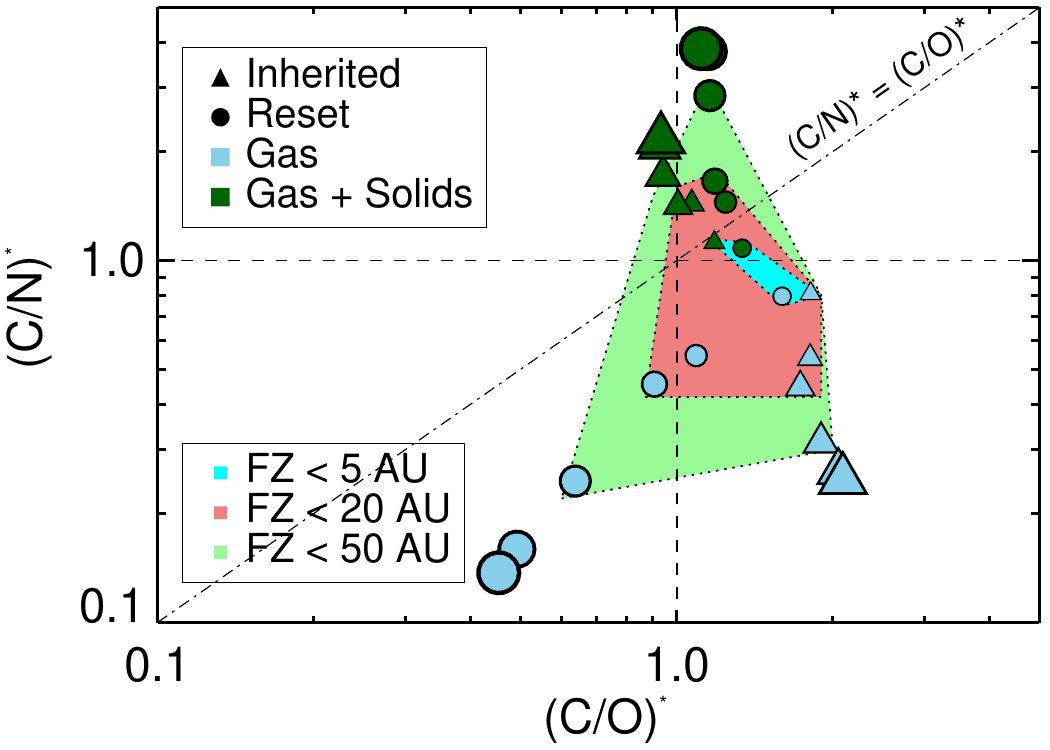}
    \caption{Composition diversity traced by the C/O and C/N elemental ratio of planets starting their formation in different zones and chemical environments of a modeled PPD. Adapted from \citet{Pacetti2022}.}
    \label{fig:PlanetComposition}
\end{figure}

On the other hand, valuable indirect constraints on the chemical compositions of PPDs can be provided by exoplanetary atmospheres \citep{Turrini2023}. Planets gathering their mass from the gas, ices, pebbles and planetesimals that constitute the disk. These building blocks contribute in different proportions to the planetary mass, and, depending on the protoplanet initial position, migration, and formation path, 
leave a different chemical signature in planet bulk composition \citep{Turrini2021, Pacetti2022}. Fig.\ref{fig:PlanetComposition} shows the \textbf{expected} diversity in composition of Jupiter-like exoplanets derived from the modelling of different formation zones (FZ), different initial disk composition, i.e. fully inherited from molecular cloud or fully reset, and two extreme modes of planet formation, through the accretion of only gas-phase volatiles or with the contribution also of solids and ices, carried by planetesimals. 

The key questions that are being tackled observationally and theoretically are:
\begin{itemize}
    \item What determines the chemical composition of disks? How much material retains its inherited composition? How relevant is the chemical reset?
    \item How much does the material composition change during the formation of planets?  How relevant are the dynamical processes in shaping the chemical composition?
    \item What are the reasons behind the large variety of disk chemical compositions observed?
\end{itemize}

\vspace{-0.5cm}
\section{Numerical modeling of protoplanetary disk thermochemistry}
\label{modelling}
Modeling PPDs represents a significant computational effort. The temporal scales and the spatial variations and gradients of temperature, density, and radiation, require not only the inclusion of a different number of physical processes but also the consideration of their close interplay. Despite simplified models being invaluable tools to explore specific regions, for example, the midplane (e.g., \citealt{Pacetti2022}), or specific configurations, like equilibrium chemical abundances of critical species (e.g., \citealt{Cleeves2015}), global modeling requires taking into account the coupling of chemistry, hydrodynamics, magnetic fields, thermal processes, radiative transfer, and dust evolution.

From the modeler's point of view, the disk can be divided into three main regions (cf., e.g., \citealt{Henning2013}): (i) the high-density cold midplane, where the stellar radiation plays a minor role, and where neutral-neutral and surface chemistry dominates; (ii) An upper low-density hot region, usually called ``atmosphere'' dominated by the presence of FUV and X-ray radiation and ion chemistry, where the temperature is controlled by atomic cooling and photoelectric and photochemical heating; (iii) An intermediate region that represents the transition between the aforementioned optically thick and thin regions, where photevaporative winds are launched, and where the thermochemical and (magneto-)hydrodynamical temporal scales become comparable. There, modeling the finer details of molecular chemistry (reactive and collisional) becomes crucial (e.g., \citealt{Sellek2022}).

Since the exact modeling of these regions depends on the specific scientific question, several numerical models have been proposed through the years. Simulations that include chemical evolution are divided into three main categories: (i) Static thermal and physical structure and variable chemical evolution (e.g., \citealt{Cleeves2015}); (ii) Evolving thermal and chemical structure, but no hydrodynamics (e.g., \citealt{Woitke2009}); (iii) Including (magneto-)hydrodynamics, temperature, and chemical evolution coupled and possibly consistent (e.g., \citealt{Gressel2020}).

In this classification, chemistry is a crucial component \citep{Oeberg2023}. For this reason, depending on the molecule of interest, these models require tuned or optimized chemical networks, representing a challenging computational cost (e.g., isotope chemistry) or limited data availability (e.g., ice chemistry or high-temperature chemistry). In this respect, including one or more ingredients among ices, X-ray chemistry, isotopologue chemistry, three-body reactions, excited H$_2$ chemistry, PAH chemistry, charged dust, and chemical mixing due to fluid advection, represents a timely challenge for planet formation and interpreting observational data. This complexity produces a number of chemical networks, rather than a “standard” one, that further increases the modelers' effort.
Analogously, computing the thermal structure of the disk is crucial because it determines the chemical evolution, the prediction of the observable quantities, but also the changes in the (magneto-)hydrodynamical quantities (e.g., \citealt{Woitke2009,Gressel2020}). It is also challenging since the thermal history depends not only on surface and dust chemistry, but also on the radiative and cosmic rays transfer. In addition, thermal processes might not be well constrained, especially when non-standard radiation fields are included (e.g., X-rays, \citealt{Ercolano2017}), or when the detailed parameters of the model are not constrained (e.g., photoelectric heating and PAH, \citealt{Kamp2011}).

Chemistry, and especially its interaction with the dust, has a significant impact on non-ideal MHD mechanisms, that are controlled by the actual ionization fraction and on the momentum transfer between ions and neutral \citep{Pinto2008}. In this case, the specific grain size distribution has a profound impact on the diffusion of the magnetic fields. Therefore, non-ideal MHD is tightly connected with the different grain properties in different regions of the disk \citep{Zhao2018}.
In addition, to ``standard'' processes, there are additional ingredients, that are rarely or never, included, but that might be pivotal for understanding observations. An example is cosmic rays produced by stellar flares, which might play a key role in interpreting the observations of the inner disk regions \citep{Brunn2023}.

This limited overview indicates that disk modeling requires a large number of interconnected processes. However, including all of them, is an overwhelming challenge, that requires an exceptional coding and computational cost. To alleviate this effort, modelers need to know what the \emph{non-relevant} processes are and what the key parameters are. To this aim, mapping the input parameters and the importance of each physical process against the observables through sensitivity analysis is a key tool. In parallel, machine learning (especially the physics-informed and interpretable one) is showing promising results in reducing the computational costs of these simulations \citep{Grassi2022}.  
\vspace{0.1cm}

\noindent {\bf{Affiliations}}\par
$^{6}$ INAF - IAPS, via Fosso del Cavaliere, 100, I-00133 Roma, Italy \par
$^{7}$ Alma Mater Studiorum Università di Bologna, Dipartimento di Fisica e Astronomia (DIFA), Via Gobetti 93/2, I-40129, Bologna, Italy\par
$^{8}$INAF-Osservatorio Astrofisico di Arcetri, Largo E. Fermi 5, I-50125, Firenze, Italy\par

\begin{acknowledgements}
The authors kindly thank A. Houge for drawing Figure~1 and the anonymous reviewer for providing constructive comments. This work was partly supported by European Union’s Horizon 2020 research and innovation program and the European Research Council via the ERC Synergy Grant ``ECOGAL'' (project ID 855130). E.S. acknowledges the support of the ASI-INAF agreement 2021-5-HH.0 and its addendum, as well as contribution from PRIN INAF 2019 through the project “HOT-ATMOS”.
\end{acknowledgements}

\bibliographystyle{aa}
% \bibliography{bibliography}

\end{document}